# What is a physical measure of spatial inhomogeneity comparable to the mathematical approach?

Z. Garncarek[1] and R. Piasecki[2, a]

[1] Institute of Mathematics, University of Opole, Oleska 48, 45052 Opole, Poland
[2] Institute of Chemistry, University of Opole, Oleska 48, 45052 Opole, Poland

**Abstract.** A linear transformation $f(S)$ of configurational entropy with length scale dependent coefficients as a measure of spatial inhomogeneity is considered. When a final pattern is formed with periodically repeated initial arrangement of point objects the value of the measure is conserved. This property allows for computation of the measure at every length scale. Its remarkable sensitivity to the deviation (per cell) from a possible maximally uniform object distribution for the length scale considered is comparable to behaviour of strictly mathematical measure $h$ introduced by Garncarek et al. in [2]. Computer generated object distributions reveal a correlation between the two measures at a given length scale for *all* configurations as well as at *all* length scales for a given configuration. Some examples of complementary behaviour of the two measures are indicated.

## 1 Introduction

Over the last couple of years the establishments of reasonably simple tools for the quantitative characterization of correlated random microstructures have been of increasing significance in various areas of science. Such tools are important for the investigation of correlation between the macroscopic properties and microstructure attributes of the medium. Recently, a comprehensive review devoted to that point for porous structures has been given [1]. Let us indicate some other examples of media with a complex spatial distribution of objects such as metallic islands, carbon black or grains. A simple mathematical method [2] for the quantitative characterization of spatial inhomogeneity of point object distribution was employed to analyze transmission electron micrographs of thin metallic films [3, 4] and polymer / carbon black composites [5]. Also other methods used in mathematical statistics were considered. Recently, to assess dislocation distributions, correlation and Poisson statistics methods have been used [6]. One more variant of this practical approach to homogeneity characterization of binary grain mixture by comparison of actual variance of point markers of each phase with the variance of ideal homogeneous mixture was applied in [7]. From a geometric statistics viewpoint, other characteristics for point fields can be found in [8].

On the other hand a recent study [9] has proposed a quantitative characterization of morphological features of a material system based on a normalized information entropy function. This measure is more general than the "local porosity entropy" [10] and the "configuration entropy" [11]. The last two entropy concepts were found to be rigorously connected [12]. Independent suggestions have been made by Garncarek, Rudnicki, Flin and others [13] about a possible connection between the entropy and Garncarek's mathematical approach. We have found a useful physical measure that is correlated to the mathematical one as was mentioned briefly in [5].

The present paper considers a linear transformation $f(S) = a \cdot S + b$ of the configurational entropy $S$ with length-scale-dependent coefficients. Our proposition, specified in the next section, differs from the other entropic approaches mentioned above. For every length scale, it provides a simple method of quantitative characterization of the most common form of spatial inhomogeneities in which fine particles or clusters of them are present in an otherwise homogeneous medium. The physical significance of the measure $f(S)$ is demonstrated on examples of computer simulated distributions of Gaussian particle clusters (more or less compact). When a distribution of a given number of objects in a number of cells is examined for all length scales, the peaks of $f(S)$ occur at those length scales for which the clustering process is relatively strong. On the other hand, the deep minima in $f(S)$ occur at length scales for which partially regular arrangements appear. It is shown that such length scales reflect the periodicity of the microstructure. Also, we show on simple examples that there is a correlation between the physical measure $f(S)$ and Garncarek's mathematical approach.

## 2 Configurational entropy measure

Consider as a starting point a model used by Frieden [14], also employed in [15, 16], where maximum entropy image processing has been discussed. Let a square image of side length $L$ in pixels be treated as a set of $\chi$ boxes (lattice cells) in which are randomly distributed $n$ balls (considered as point objects) representing small grains of silver on a photographic negative. For each scale $k = (L^2 / \chi)^{1/2}$ with a given distribution $(n_1, ..., n_i, ..., n_\chi)$, *i.e.* having fixed numbers $n_i$ of point objects in $i$th cell and fulfilling constraint $n_1 + n_2 + ... + n_\chi = n$, one can associate a configurational entropy $S = k_B \ln \Omega$, where the Boltzmann constant will be set to $k_B = 1$ for convenience. Note that $k$ is equal to the lattice cell side-length. The degeneracy factor $\Omega$, that is, the number of different ways of generating the fixed distribution, is given by

$$\Omega \equiv \frac{n!}{n_1! n_2! \ldots n_\chi!} \ . \tag{1}$$

Note that Stirling's approximation is not used here since we do not assume that all the $n_i$ are large numbers. The highest possible value of configurational entropy at a given length scale is related to the most uniformly distributed object configuration with a degeneracy factor $\Omega_{max}$ and is denoted by $S_{max} = \ln \Omega_{max}$. One can find $\Omega_{max}$ by fixing $n_0 + 1$ and $n_0$ point objects in $r_0$ and $\chi - r_0$ cells, where $r_0 = n \ [\mathrm{mod} \ \chi]$ and $n_0 = (n - r_0) / \chi$. To evaluate for each length scale $k$ the deviation of the actual configuration from the most spatially uniform distribution it is natural to consider the difference $S_{max} - S$. However, many computer generated distributions show that, after averaging $S_{max} - S$ over the number of cells $\chi$, a high sensitivity to spatial arrangements at each length scale $k$ is revealed. Therefore, the final form of the measure is chosen to be

$$f(S) = \frac{S_{\max} - S}{\chi}$$

$$\equiv -\frac{r_0}{\chi} \ln (n_0 + 1) + \frac{1}{\chi} \sum_{i=1}^{\chi} \ln \left( \frac{n_i!}{n_0!} \right). \qquad (2)$$

The spectrum of $f(S)$ values satisfies a number of simple properties illustrated in the next section.

1. The lowest value equals to 0. It is always reached at the boundary length scales $k = 1$ and $k = L$. Otherwise, the value $f(S) = 0$ characterizes the most spatially uniform configuration, when for each pair $i \neq j$ we have $|n_i - n_j| \leq 1$, (see, for example possible at a given scale configuration 1 in Tab. 1 or the right grids in Figs. 5b and 5c). For an ideally homogeneous spatial distribution at a given length scale, we have $r_0 = 0$ and each $n_i = n_0 \equiv n / \chi$.

2. The highest value for a given length scale is reached when all objects are placed in one lattice cell. This case corresponds to the strongest deviation per cell from a possible maximally uniform distribution for the length scale considered (see, for example configuration 18 in Tab. 1).

3. The first peak of $f(S)$ corresponds to clustering of objects (see, for example the left grids in Figs. 5b and 5c). However, the subdivision of bigger clusters also contributes to such deviations. Finding a reliable interpretation of the other maxima needs a more detailed investigation for each case, for example, the shifted initial arrangement.

4. The deep minima in $f(S)$ describe partially regular arrangements of objects. The sequential positions of such minima reflect periodicity of the whole microstructure. For the simple case of one deep minimum, see Figures 3b and 4b, where the periodicity is shown.

5. This property allows us to calculate the exact value of $f(S)$ at every length scale: when the final pattern of size $mL \times mL$ (where $m$ is a natural number) is formed by periodical repetition of an initial arrangement of size $L \times L$ then the value of the measure is conserved. We can distinguish between length scales that are commensurate and incommensurate with the side length $L$. If $L \, [\text{mod } k] = 0$ then $f(S(k, L)) = f(S(k, mL))$. For the latter case we have $L \, [\text{mod } k] \neq 0$ and we must find a natural number $m'$ fulfilling the condition $(m'L) \, [\text{mod } k] = 0$. Then we define $f(S(k, L)) = f(S(k, m'L))$.

Interestingly, a suggested connection between $f(S)$ and $h$ appears. For easy reference we recall here the definition of mathematical measure $h$ (the proof can be found in the monograph [17] and was summarized in the Appendix in [5])

$$h = \frac{\mu}{E(\mu)} \equiv -\frac{n}{\chi - 1} + \frac{\chi}{n(\chi - 1)} \sum_{i=1}^{\chi} n_i^2, \qquad (3)$$

where $\mu \equiv \Sigma_{i=1}^{\chi}(n_i - n / \chi)^2$, $E(\mu) \equiv n(\chi - 1) / \chi$ is the expectation value of the random variable $\mu$, $n$ and $\chi$ describe as usual the number of point objects and lattice cells. The possible values of the measure $h$ belong to $[0, n]$. When $h = 0$, a distribution is perfectly homogeneous, for $h = 1$ it is random and for $h = n$ the maximum value of its spatial inhomogeneity is reached.

## 3 Examples of correlation between the two measures and results

To illustrate the similarity of the behaviour of $f(S)$ and $h$, we restrict ourselves to a few simple examples of computer generated configurations. We distinguish the two situations:

I. for a fixed length-scale, the values of the two measures are considered for all possible point object configurations;

II. for a given pixel configuration, the set of pairs $(h, f(S))$ is considered at all ∕ some length scales.

Let us consider the two examples belonging to group I. The first example is for $n = 9$ point objects and $\chi = 4$. The corresponding values of the two measures for the representative configurations $(n_1, n_2, n_3, n_4)$ are given in Table 1.

**Table 1.** Comparison between the values of $f(S)$ and $h$ calculated for representative configurations.

| No. | Configuration | | | | $f(S)$ | $h$ |
|---|---|---|---|---|---|---|
| 1 | 3 | 2 | 2 | 2 | 0.00 | 0.11 |
| 2 | 3 | 3 | 2 | 1 | 0.10 | 0.41 |
| 3 | 4 | 2 | 2 | 1 | 0.17 | 0.70 |
| 4 | 4 | 3 | 1 | 1 | 0.28 | 1.00 |
| 5 | 3 | 3 | 3 | 0 | 0.38 | 1.00 |
| 6 | 5 | 2 | 1 | 1 | 0.40 | 1.59 |
| 7 | 4 | 3 | 2 | 0 | 0.45 | 1.30 |
| 8 | 5 | 2 | 2 | 0 | 0.58 | 1.89 |
| 9 | 4 | 4 | 1 | 0 | 0.62 | 1.89 |
| 10 | 5 | 3 | 1 | 0 | 0.68 | 2.19 |
| 11 | 6 | 1 | 1 | 1 | 0.68 | 2.78 |
| 12 | 6 | 2 | 1 | 0 | 0.85 | 3.07 |
| 13 | 5 | 4 | 0 | 0 | 1.02 | 3.07 |
| 14 | 6 | 3 | 0 | 0 | 1.13 | 3.67 |
| 15 | 7 | 1 | 1 | 0 | 1.16 | 4.56 |
| 16 | 7 | 2 | 0 | 0 | 1.34 | 4.85 |
| 17 | 8 | 1 | 0 | 0 | 1.68 | 6.63 |
| 18 | 9 | 0 | 0 | 0 | 2.23 | 9.00 |

The two sets of values of $h$ and $f(S)$ for representative configurations are shown in Figure 1. The association between the values graphically represented is evident (linear correlation coefficient $r = 0.9933$). The pairs of configurations distinguished by $f(S)$ but not by $h$, that is 4 and 5, 8 and 9, 12 and 13 (a reverse situation appears for the pair 10 and 11) show that different sets of the occupations numbers $n_i$ may produce the same value for one of the measures. However, from the mathematical statistics viewpoint those cases do not contradict the association observed but only decrease the correlation. On the other hand a functional relation $h = G_1(f(S))$ and $f(S) = G_2(h)$ is forbidden.

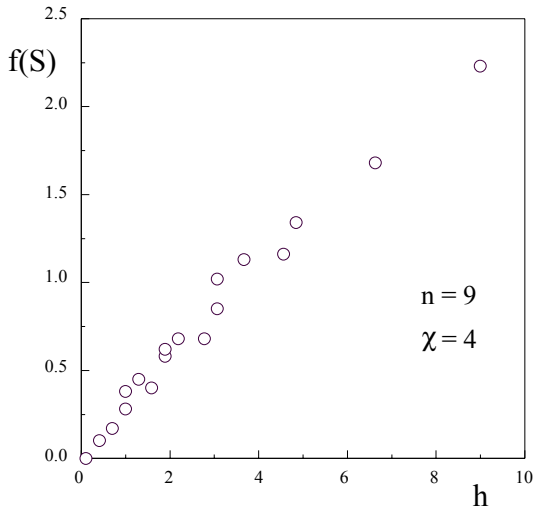
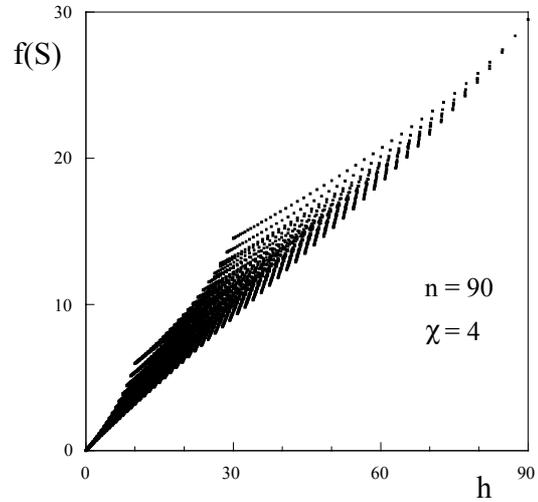

Fig. 1. The values of the two measures $h$ and $f(S) = (S_{max} - S) / \chi$ computed for all representative configurations of $n = 9$ point objects placed in $\chi = 4$ square cells are presented. The strong correlation between the two measures showed by locations of points $(h, f(S))$ is confirmed by a linear correlation coefficient $r = 0.9933$.

Fig. 2. Same as in Figure 1, with $n = 90$ point objects placed in $\chi = 4$ square cells. A linear correlation coefficient is equal to $r = 0.9883$. This kind of correlation is particularly strong for distributions approaching the two boundary configurations: the perfectly homogeneous and the most inhomogeneous.

The second example describes $n = 90$ point objects placed again in $\chi = 4$ cells. The two sets of values of $h$ and $f(S)$ for representative configurations are shown in Figure 2. The association between the values graphically represented is easily seen (linear correlation coefficient $r = 0.9883$). Such a correlation is the strongest for the highly regular configurations as well as the highly inhomogeneous distributions.

Now, let us consider the following examples belonging to group II. The two representative initial arrangements of 24 clusters (6 per quadrant) each composed of 3 black pixels, were computer generated in a square of $20 \times 20$ pixels. Figures 3a and 4a show the picked Gaussian clusters for which black pixels were distributed with a standard deviation $\sigma = 1.0$ and $\sigma = 1.5$, respectively. In Figures 3b and 4b the behaviour of the two measures at *all* length scales $k$ is presented. To obtain results for all length scales $k$ we employed the property 5 satisfied by $f(S)$. The two initial arrangements from Figures 3a and 4a were periodically repeated until the final pattern was obtained at the length scale considered. Such approach was also suggested in [9] to ameliorate the problem of large deviations of normalized information entropy measure, where the information entropy for the random particle configuration is subtracted from the actual information entropy. As concerns the measure $h$, its variability along the increasing size of the final patterns is within acceptable range (see Tab. 2 for the first sequential length scales) and does not change the qualitative picture for examples considered.

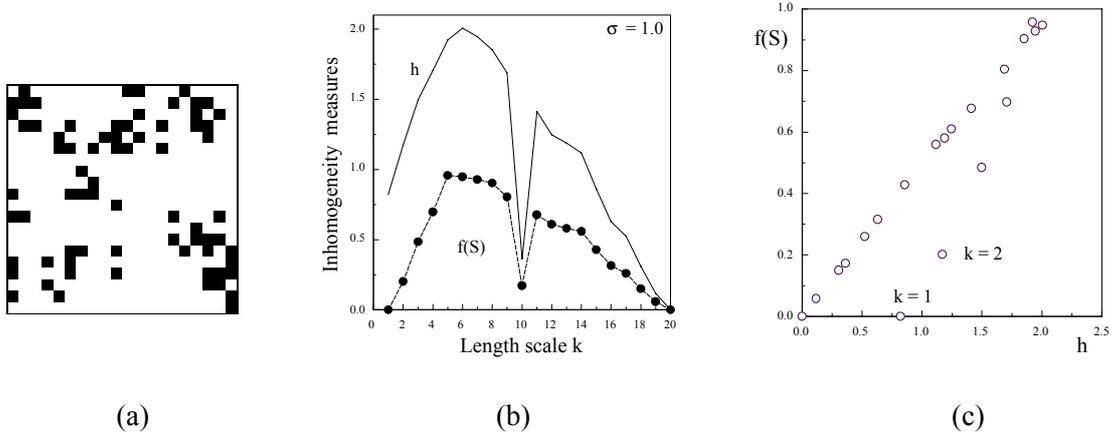

(a)                          (b)                          (c)

Fig. 3. An example of computer generated distribution and corresponding numerical results for all length scales. (a) Random initial arrangement of 24 Gaussian clusters (6 per quadrant), each composed of 3 pixels, in a square of $20 \times 20$ with a standard deviation $\sigma = 1.0$. The fraction of black pixels equals to 0.180. (b) The inhomogeneity degree measure $h$, solid line, and the measure $f(S)$, dashed line, as a function of the length scale $k$. (c) The strong correlation between the two measures showed by locations of points $(h, f(S))$ is confirmed by a linear correlation coefficient $r = 0.9643$.

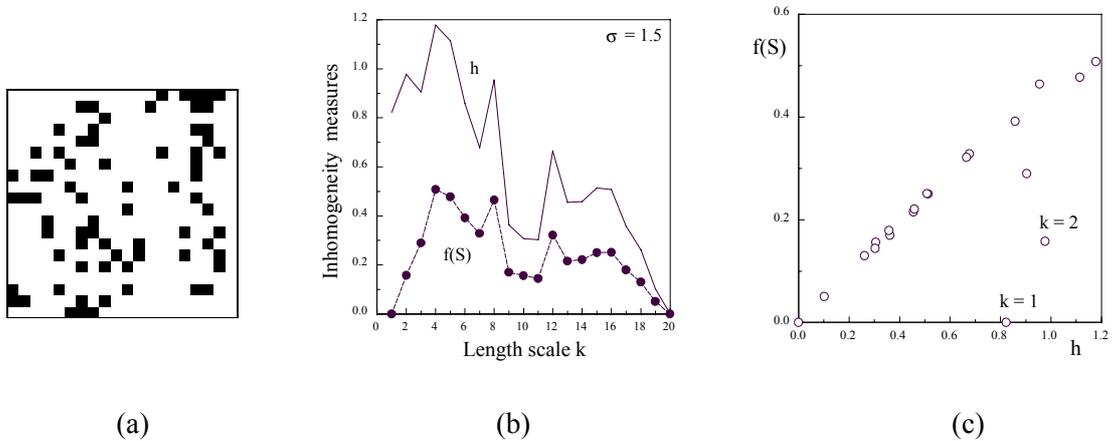

(a)                          (b)                          (c)

Fig. 4. An example of computer generated distribution and corresponding numerical results for all length scales. Same as in Figure 3, with $\sigma = 1.5$ and $r = 0.8968$.

For the measure $h$ in Figures 3b and 4b all the diagonal values for each of the two final patterns considered were chosen from a general table, i.e. created in the same way as Table 2 up to the grid $400 \times 400$ and $k = 20$. The similar behaviour of the measures $h$ and $f(S)$ is still observed. This observation is also confirmed by the easily seen associations in Figures 3c and 4c (linear correlation coefficient $r = 0.9643$ and $r = 0.8968$, respectively).

**Table 2.** The values of measure $h$ calculated for final periodic system of different sizes generated by the two initial arrangements.

| Grid | $k = 1$ | $k = 2$ | $k = 3$ | $k = 4$ | $k = 5$ |
|------|---------|---------|---------|---------|---------|
| For the initial arrangement (Fig. 3a) | | | | | |
| $20 \times 20$ | 0.8221 | 1.1807 | --- | 1.7743 | 2.0444 |
| $40 \times 40$ | 0.8205 | 1.1718 | --- | 1.7205 | 1.9471 |
| $60 \times 60$ | 0.8202 | 1.1702 | 1.4979 | 1.7109 | 1.9301 |
| $80 \times 80$ | 0.8201 | 1.1696 | --- | 1.7076 | 1.9242 |
| $100 \times 100$ | 0.8201 | 1.1694 | --- | 1.7061 | 1.9215 |
| For the initial arrangement (Fig. 4a) | | | | | |
| $20 \times 20$ | 0.8221 | 0.9843 | --- | 1.2245 | 1.1852 |
| $40 \times 40$ | 0.8205 | 0.9769 | --- | 1.1874 | 1.1287 |
| $60 \times 60$ | 0.8202 | 0.9755 | 0.9039 | 1.1808 | 1.1189 |
| $80 \times 80$ | 0.8201 | 0.9751 | --- | 1.1785 | 1.1155 |
| $100 \times 100$ | 0.8201 | 0.9748 | --- | 1.1774 | 1.1139 |

For comparison purposes, in Figure 5a the values of $f(S)$ for the two cases with $\sigma = 1.0$ and $\sigma = 1.5$ are presented.

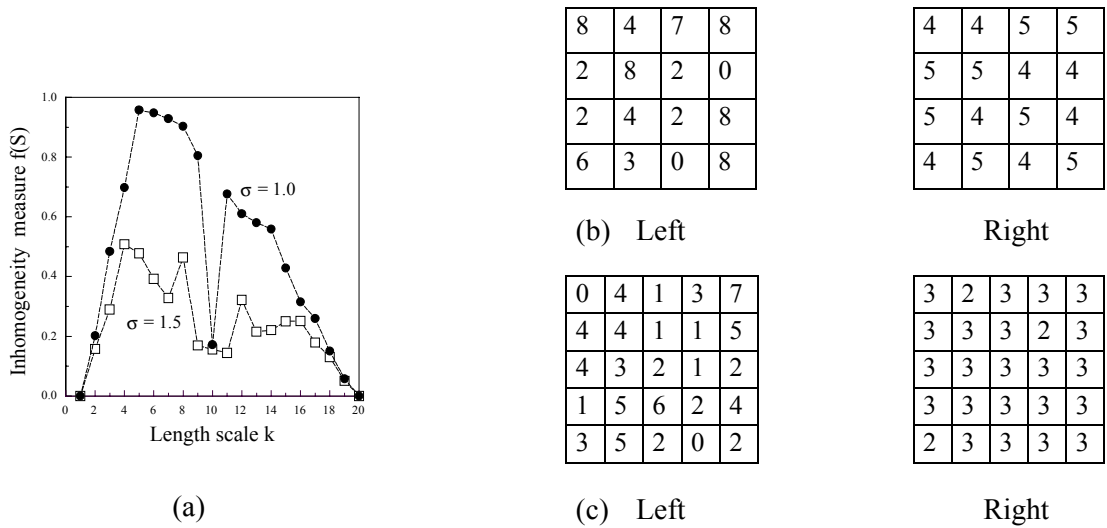

(a)

(b)   Left                          Right

(c)   Left                          Right

Fig. 5. The comparison of the measure $f(S)$ values calculated for computer generated two initial arrangements showed in Figures 3a and 4a. (a) The measure $f(S; \sigma = 1.0)$, filled circles, and $f(S; \sigma = 1.5)$, open squares, as a function of the length scale $k$. (b) The left grid with object occupation numbers (see Fig. 3a) is referred to the peak of $f(S; \sigma = 1.0)$ at $k = 5$ while the right grid to a possible maximally uniform distribution. (c) same as previously, with $\sigma = 1.5$, $k = 4$ and Figure 4a.

As expected, we observe that $f(S; \sigma = 1.0) > f(S; \sigma = 1.5)$ for most of length scales. The characteristic deep minima of both curves, the narrow one at $k = 10$ and the wider one for $9 \leq k \leq 11$ correspond to the intended periodicity caused by the procedure of picking of clusters, namely 6 per quadrant. This feature accords with the behaviour of normalized information entropy function calculated for specific periodic configurations of particles [9] (*cf.* Fig. 7). On the other hand the peaks of $f(S; \sigma = 1.0)$ at $k = 5$ and of $f(S; \sigma = 1.5)$ at $k = 4$ correspond to the relatively strongest deviations per cell from a corresponding maximally uniform distribution. The details of this behaviour are illustrated in Figure 5b for $k = 5$ and Figure 5c for $k = 4$, where the grids with object occupation numbers referring to the peaks and possible maximally uniform distributions are presented.

The two representative random distributions of 600 clusters, each composed of 12 black pixels, were computer generated in a square of $240 \times 240$ pixels. Figure 6a shows the picked clusters, in which black pixels were Gaussian distributed with a standard deviation $\sigma = 2.0$, while Figure 7a illustrates the case with $\sigma = 4.0$. The behaviour of the two measures is presented in Figures 6b and 7b. To save computation time, the calculations were performed for commensurate length scales only. Again, the similar behaviour of $h$ and $f(S)$ curves appear and is confirmed in Figures 6c and 7c (linear correlation coefficient $r = 0.9947$ and $r = 0.9966$, respectively).

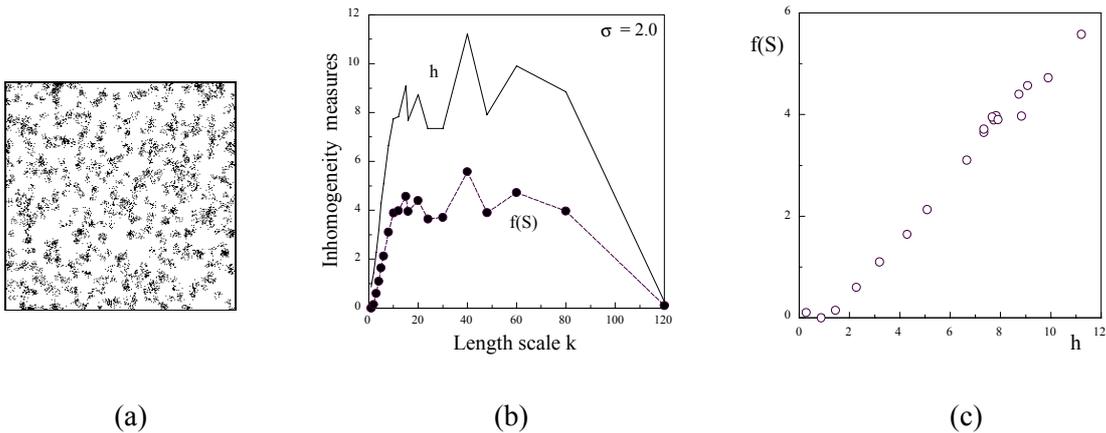

(a)                     (b)                                   (c)

Fig. 6. An example of computer generated distribution and corresponding numerical results for commensurate length scales. (a) Random initial arrangement of 600 Gaussian clusters (150 per quadrant), each composed of 12 pixels, in a square of $240 \times 240$ with a standard deviation $\sigma = 2.0$. The fraction of black pixels equals to 0.125. b) The inhomogeneity degree measure $h$, solid line, and the measure $f(S)$, dashed line, as a function of the length scale $k$. (c) The strong correlation between the two measures showed by locations of points $(h, f(S))$ is confirmed by a linear correlation coefficient $r = 0.9947$.

It is worth nothing that in this paper the measures for pairs of object distributions with identical numbers of objects, i.e. $n = 72$ (see Figs. 3a and 4a, also Fig. 5a) and $n = 7200$ (see Figs. 6a and 7a) were compared. The reason is that the measures $h$ and $f(S)$ cannot be safely used to make a comparative analysis of spatial distributions with a different number of objects. In such a case, the standardized version of $h$ previously employed in [3,4 and 5] and fully described in a monograph [17] is an appropriate measure.

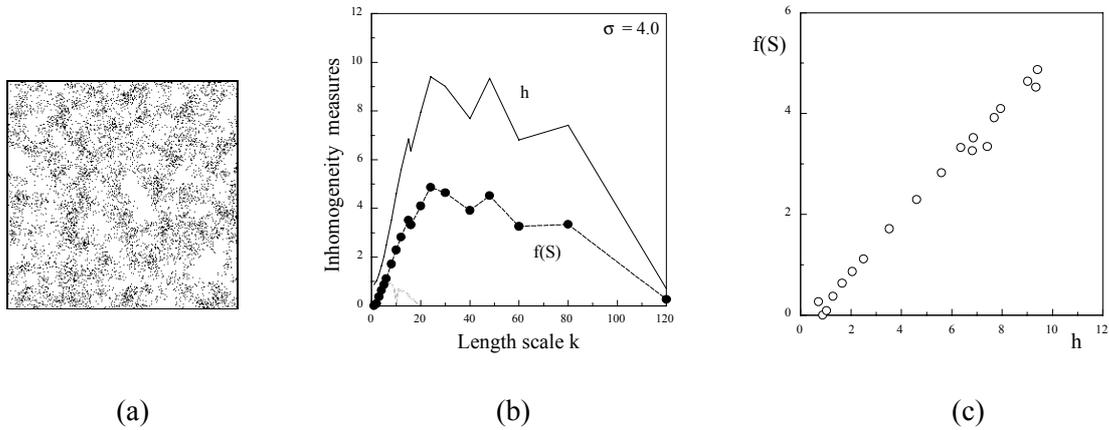

(a)              (b)                 (c)

Fig. 7. An example of computer generated distribution and corresponding numerical results for commensurate length scales. Same as in Figure 6, with $\sigma = 4.0$ and $r = 0.9966$.

In conclusion, the present work shows that the two approaches, one originating in physics and second based on mathematics, lead to strongly correlated results. It was found that $f(S)$ and $h$ are correlated at the same length scale for all representative configurations in Figures 1 and 2 (group I examples) and at all length scales for the same configuration as shown in Figures 3c and 4c for $20 \times 20$ initial arrangements periodically repeated as well as Figures 6c and 7c for $240 \times 240$ binary images (group II examples). For all cases, the two measure values, graphically represented in Figures 1, 2, 3c, 4c, 6c and 7c, are strongly correlated and this observation is supported by the appropriate linear correlation coefficients $r = 0.9933$, $0.9883$, $0.9643$, $0.8968$, $0.9947$ and $0.9966$.

Obviously, that does not mean that the two measures always provide the same information. For particular cases the behaviour of the two measures indicating a weakening of their correlation cannot be ignored. For group I, arrangements indistinguishable by one of the measures but distinguishable by the second can be observed (see Tab. 1). Similarly, for group II see the points corresponding to the scales $k = 1$ and $k = 2$ in Figures 3c and 4c. Thus, the two measures rather complement each other for such configurations and scales.

We have shown that the physical measure $f(S)$ provides a new useful and alternative method of evaluation of the spatial distribution of objects. For applications, the complementary behaviour of the two measures at some cases can be also relevant.